\documentclass[twocolumn]{aastex702}

\usepackage{amsmath}
\graphicspath{{figures/}}

\newcommand{\dmunit}{\mathrm{pc\,cm^{-3}}}
\newcommand{\kms}{\mathrm{km\,s^{-1}}}
\newcommand{\sigpos}{\sigma_{\rm pos}}
\newcommand{\fhost}{f_{\rm host}}
\newcommand{\ADM}{A_{\rm DM}}
\newcommand{\thEobs}{\theta_{E,\rm obs}}
\newcommand{\DDM}{|\Delta\mathrm{DM}|}
\newcommand{\sigrep}{\sigma_{\rm rep}}
\newcommand{\DDMlens}{|\Delta\mathrm{DM}_{\rm lens}|}

\shorttitle{Identifying strongly lensed FRBs}
\shortauthors{Qi et al.}

\begin{document}

\title{Identifying Strongly Lensed Fast Radio Bursts with Optical Galaxy Catalogs}

\author[orcid=0000-0001-7492-874X,gname=Jing-Zhao,sname=Qi]{Jing-Zhao Qi}
\affiliation{Department of Physics, College of Sciences, Northeastern University, Shenyang 110819, China}
\email{qijingzhao@mail.neu.edu.cn}

\author[orcid=0009-0004-5607-9181,gname=Ji-Guo,sname=Zhang]{Ji-Guo Zhang}
\affiliation{Department of Physics, College of Sciences, Northeastern University, Shenyang 110819, China}
\email{zhangjiguo@stumail.neu.edu.cn}

\author[orcid=0000-0002-6029-1933,gname=Xin,sname=Zhang]{Xin Zhang}
\affiliation{Department of Physics, College of Sciences, Northeastern University, Shenyang 110819, China}
\affiliation{Liaoning Key Laboratory of Cosmology and Astrophysics, Northeastern University, Shenyang 110819, China}
\affiliation{MOE Key Laboratory of Data Analytics and Optimization for Smart Industry, Northeastern University, Shenyang 110819, China}
\affiliation{National Frontiers Science Center for Industrial Intelligence and Systems Optimization, Northeastern University, Shenyang 110819, China}
\email[show]{zhangxin@mail.neu.edu.cn}
\correspondingauthor{Xin Zhang}

\begin{abstract}
Strongly lensed fast radio bursts (FRBs) would provide millisecond-precision time delays and thus a measurement of the Hubble constant independent of both the distance ladder and the early Universe. Using them first requires identifying lensed image pairs among tens of thousands of bursts, where burst pairs of unlensed repeaters, alike in position, arrival time, and dispersion measure (DM), outnumber them by more than an order of magnitude. Radio data alone cannot separate them reliably, because a DM-difference cut works only if its threshold matches DM variations of repeaters and lenses that are unknown in advance. We propose a new method that identifies strongly lensed FRBs with optical galaxy catalogs. It tests whether a foreground galaxy could have produced the pair, with position, redshift, and Einstein radius matching the pair and a maximum delay exceeding the observed interval, whereas a repeater's host lies at the source redshift and cannot. In simulations of a five-year all-sky survey, the method raises the fraction of genuine lensed pairs among candidates from 0.35 to 0.97--0.99 at a fixed DM cut, and from 0.02 to 0.46--0.93 for three times larger repeater DM variations, if repeater hosts are largely absent from the catalog. An optimistic catalog of galaxy--galaxy strong lenses keeps it near 0.8 without any DM-based rejection, even when the images have no DM difference. Optical catalogs thus provide an effective route to identifying strongly lensed FRBs and to using them for time-delay cosmology.
\end{abstract}


\section{Introduction} \label{sec:intro}

The local measurement of the Hubble constant $H_0$ and the value inferred from the cosmic microwave background in the standard cosmological model differ at about the $5\sigma$ level \citep{Riess2022H0,Planck2018}. Time delays in strong gravitational lenses measure the time-delay distance and provide a route to $H_0$ that is independent of the distance ladder \citep{TreuMarshall2016,Wang:2021kxc,Qi:2022sxm,Song:2025ddm}. For lensed quasars, a decade of monitoring yields delays with percent-level precision \citep{Tewes:2012zz,HOLiCOW:2019oog}. Fast radio bursts (FRBs) last about a millisecond, so the arrival time of a single burst is known to better than a millisecond, and for a delay of 30 days the relative precision is about $4\times10^{-10}$. The lens model, line-of-sight structure, and redshifts still limit $H_0$, but strongly lensed FRBs remove the error of the delay itself \citep{Li:2017mek,Wucknitz:2020spz,Zhang:2024rra} and complement cosmology with the DMs of unlensed FRBs \citep{Zhao:2020ole,Zhao:2022yiv,Zhang:2023gye}, which combine with lensing time delays and other late-universe probes \citep{Wu:2022dgy,Wu:2026mox}.

To exploit strongly lensed FRBs, the image pairs must first be identified in a radio catalog. The next generation of wide-field radio interferometers will record on the order of ten thousand FRBs per year, among which a small number of galaxy-scale strong-lensing events are expected \citep{Connor:2022bwl}. The two images are about one arcsecond apart, comparable to the localization error of these arrays, and arrive days to months apart, so they must be found by comparing events recorded at different times. A fraction of FRB sources repeat, for example 18 of the 492 sources in the first CHIME/FRB catalog \citep{CHIMEFRB2021Catalog}, and the spectral morphology of bursts suggests that some apparently non-repeating sources also repeat \citep{Sun:2024huw,Sun:2025cdq,Sun:2026aeh}. Two bursts of an unlensed repeater are also measured about one arcsecond apart because of the localization errors, the interval between them can fall within the range of lensing delays, and their dispersion measures (DMs) are nearly equal. In a catalog of tens of thousands of events, such repeater pairs outnumber genuine image pairs by more than an order of magnitude (Section~\ref{sec:background}). To our knowledge, no galaxy-scale strongly lensed FRB has yet been identified individually, and time-delay cosmology with FRBs first requires a reliable route to such identifications.

Existing searches for lensed FRBs rely on radio data. Baseband searches for coherent echoes target microlensing and millilensing by low-mass lenses \citep{CHIMEFRB:2022xzl,Leung:2023lmq}, and catalog searches compare the dynamic spectra of nearby bursts \citep{Chang:2024hxi,Kader:2025tia}. At the catalog level, the most direct radio discriminant is the DM difference. The two light paths of a lensed FRB cross different parts of the lens galaxy, so their DMs can differ by several to tens of $\dmunit$ \citep{Er:2022lad}, whereas bursts of one repeater usually have very similar DMs. The gas content of lens galaxies is poorly known, and the DMs of repeaters vary. FRB 20190520B decreased in DM by $12.4~\dmunit$ per year over four years \citep{Niu:2025iin}, and at least three repeaters show secular DM trends \citep{Zhang:2026nbr}. Optical data have so far been used statistically, for example by cross-matching FRBs with massive galaxy clusters \citep{Sammons:2025ogt,Sammons:2026pto}, and optical lens catalogs have been proposed to localize and identify lensed gravitational waves \citep{Hannuksela:2020xor,Liu:2025gkq}. Wide-field surveys will provide hundreds of millions of galaxies and tens of thousands of galaxy--galaxy strong lenses (GGSLs) \citep{Collett:2015roa,Euclid:2025rfv,DESI:2025tvs}. Whether such catalogs can exclude repeater pairs one by one has not been assessed.

In this Letter, we propose a new method that identifies strongly lensed FRBs with optical galaxy catalogs. We show that a radio-only DM cut works only if its threshold matches the uncertain DM behavior of repeaters and lenses, and test under which conditions optical galaxy catalogs distinguish lensed image pairs from repeater pairs with less dependence on it. The optical distinction rests on the physics of lensing. Two images of one burst require a foreground galaxy whose position, redshift, and mass account for the separation and the arrival-time difference of the two events. Two bursts of a repeater have no such galaxy. Their host lies between the two measured positions, but at the source redshift, so it cannot produce a one-arcsecond separation. With simulations of a five-year all-sky survey, we compare both criteria on the same catalogs. If repeater hosts are largely absent from the general galaxy catalog, then at a fixed DM cut the optical criterion raises the fraction of genuine lensed pairs among the candidates by a factor of about three in our fiducial model, where a better-tuned DM cut does as well, and by factors of 27--54 for a three times larger repeater DM variation, at the cost of about half or more of the image pairs. An optimistic LSST-like catalog of galaxy--galaxy strong lenses works without DM-based rejection of repeaters, and its purity and recovery do not depend on the DM conditions.

\section{Repeater Background and the Basis of the Optical Criterion} \label{sec:background}

\begin{figure*}
\centering
\includegraphics[width=\textwidth]{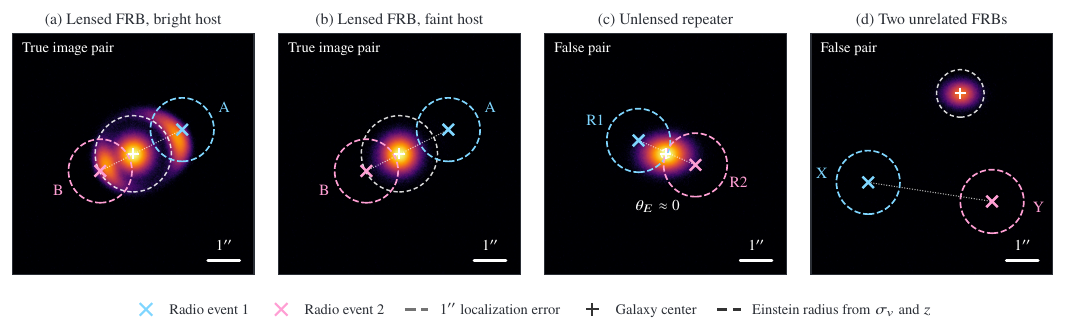}
\caption{Origins of event pairs in the catalog (schematic). (a) The two lensed images of one burst lie on opposite sides of the lens galaxy, and a bright host is stretched into arcs. (b) The same system with a faint host. (c) Two bursts of an unlensed repeater, about one arcsecond apart because of the localization errors, with the host between them. (d) Two unrelated FRBs and an unrelated galaxy. Crosses mark the radio positions, colored dashed circles the $1''$ localization error, white plus signs the galaxy centers, and white dashed circles the Einstein radius inferred from the velocity dispersion and redshift of the galaxy. This radius matches the image separation in (a) and (b), is much smaller than the separation for the host in (c), and the galaxy in (d) is not on the line joining the events.
\label{fig:cases}}
\end{figure*}

An event is a catalog record of position, arrival time, and DM. Two close events form an event pair, which is an image pair if they are the two images of one strongly lensed burst. An event pair that passes a selection is a candidate. For the optical schemes, a candidate is an event pair together with a catalog galaxy that could have lensed it, the deflector. Figure~\ref{fig:cases} illustrates the origins of event pairs.

In our fiducial simulation (Section~\ref{sec:sim}), after removing near-duplicate events (Section~\ref{sec:veto}) and pairing events within $10''$ whose DMs differ by less than $150~\dmunit$, the loosest DM condition used below (Section~\ref{sec:conditions}), a five-year realization contains on average 344.6 event pairs before any further selection, of which only 13.1 are genuine image pairs (the veto and the pairing conditions remove 2.1 of the 15.2 recoverable image pairs, those with both images detected within the survey, per realization). According to the simulation truth, 331.0 of the 331.5 false pairs come from unlensed repeaters, 0.42 are cross-burst pairs, that is, events of the same lensed repeater belonging to different bursts, and 0.07 are chance pairs of unrelated sources. The repeater background is large because every repeater with $n$ detected bursts yields $n(n-1)/2$ pairs, 7.5 on average in our model (Section~\ref{sec:sim}).

Radio positions and arrival times cannot separate the two classes reliably. For $\sigpos=1''$, the median measured separation is $2.01''$ for genuine image pairs and $1.66''$ for two measurements of one position, so the distributions overlap. True delays range from days to months, and the interval between two bursts of a repeater can take any value up to years. A time window shorter than the range of lensing delays would remove most repeater pairs in our simulation, where burst times are uniform, but real repeaters are clustered in time \citep{Opperman:2017kql}, so we do not use one. The radio reference of Section~\ref{sec:veto} keeps only a loose limit set by the lensing delays. The DM difference is then the main radio discriminant. Let $\sigrep$ be the standard deviation of the DMs of repeat bursts about the DM of the first burst and $\DDMlens$ the DM difference between lensed images, which grows with the ionized gas of the lens galaxy. A cut that removes pairs with $\DDM<T$ works only if $\sigrep\ll T\ll\DDMlens$, and both $\sigrep$ and $\DDMlens$ are poorly known (Section~\ref{sec:intro}).

An optical catalog does not indicate whether a galaxy is a host or a lens, and the FRB redshift is not measured. We therefore estimate the source redshift $\hat z_s$ from the mean DM of the pair (Section~\ref{sec:dmmodel}) and test whether a catalog galaxy can act as the lens. A genuine image pair requires a galaxy near the line joining the events, with a redshift below that of the source, an Einstein radius matching the separation, and a maximum time delay exceeding the observed one. For a repeater pair, the only galaxies at the right position are unrelated galaxies or the host itself. The search region for a deflector is a rectangle along the line joining the events, $6\sigpos$ wide and $s+6\sigpos$ long for a measured separation $s$ (Equation~\ref{eq:position}), and its mean area for false pairs is $A\approx56~\mathrm{arcsec^2}$. For early-type galaxies to $i=25$ with $\Sigma_{\rm gal}\approx2.33\times10^4~\mathrm{deg^{-2}}$, the expected number of unrelated galaxies in this region is
\begin{equation}
\bar{N}_{\rm gal}=\Sigma_{\rm gal}A\approx0.10, \label{eq:ngal}
\end{equation}
and only a fraction $P_{\rm phys}\approx0.07$ of them pass the mass pre-screen and the redshift, time-delay, and Einstein-radius conditions (Section~\ref{sec:mechanism}), so fewer than one percent of false pairs acquire an unrelated deflector. The host of a repeater does not follow this geometric estimate. The median FRB--host offset of 3.3~kpc \citep{Heintz2020Hosts} corresponds to $0.4''$ at $z\sim1$, so the host almost always lies in the search region and can only be excluded physically. Its true redshift equals the source redshift, so even if redshift errors place it below $\hat z_s$, $D_{ls}\ll D_s$ and its inferred Einstein radius is, for most hosts, far too small for a one-arcsecond separation (Section~\ref{sec:mechanism}).

\section{Method and Simulations} \label{sec:method}

\begin{figure*}
\centering
\includegraphics[width=\textwidth]{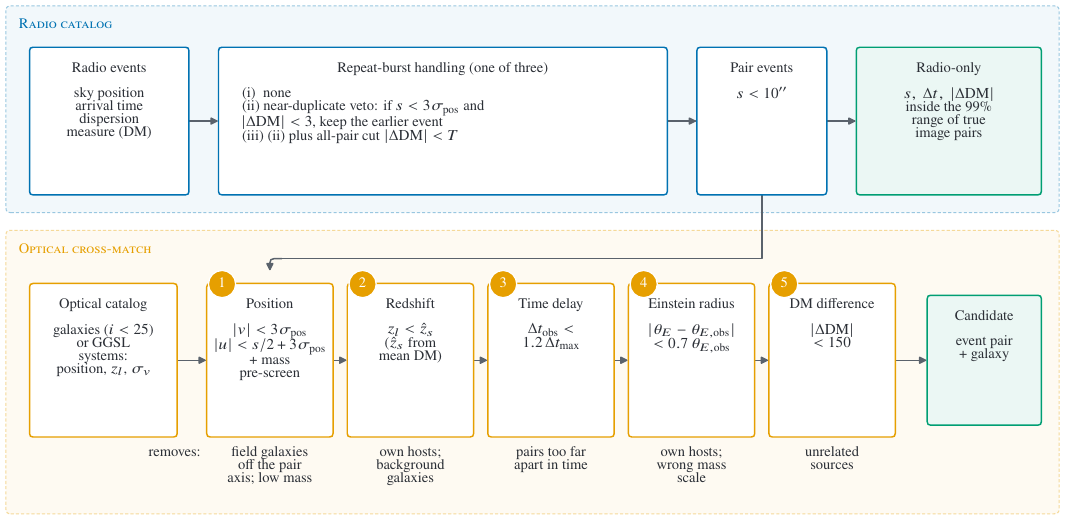}
\caption{Identification procedure. Top: processing of the radio catalog. One of three treatments of repeat bursts (Section~\ref{sec:veto}) is applied before events within $10''$ are paired, and the pairs enter either the radio-only selection, which keeps pairs inside the calibrated window, or the optical cross-match. Bottom: the optical conditions applied in turn (Section~\ref{sec:conditions}), with the background each mainly removes written below it. Every combination of an event pair and a catalog galaxy that passes all conditions is a candidate. In this work the output is scored against the simulation truth.
\label{fig:flow}}
\end{figure*}

Each realization simulates radio and optical catalogs, pairs the radio events, matches each pair to an optical catalog, and scores the output against the simulation truth. The truth is used only in this scoring step. Figure~\ref{fig:flow} shows the identification procedure, which would be applied in the same way to real catalogs, and Table~\ref{tab:model} lists the fiducial models.

\begin{deluxetable*}{lp{14.2cm}}
\tabletypesize{\footnotesize}
\tablecaption{Fiducial Models Adopted in the Simulation \label{tab:model}}
\tablehead{\colhead{Component} & \colhead{Fiducial setting}}
\startdata
Survey & All sky, five years, about $5\times10^4$ unlensed events. Detection threshold $5~\mathrm{Jy\,ms}$. Per-axis localization error $\sigpos=1''$ (also $0.3''$ to $5''$, with a pairing radius of $25''$ at $5''$) \\
FRB population & Redshift distribution proportional to the star formation rate \citep{MadauDickinson2014}, $0<z<4$. Energy function $dN/dE_\nu\propto E_\nu^{-1.5}$ for $10^{31}$ to $10^{34}~\mathrm{erg\,Hz^{-1}}$. Spectrum $E_\nu\propto\nu^{-1.5}$ \citep{Macquart2019Spectra} \\
Repeaters & 5\% of the sources. Number of bursts per source $1+\mathrm{Poisson}(3)$, with times uniformly distributed over five years. Standard deviation of the DM variation relative to the first burst $\sigrep=1~\dmunit$ (also 3) \\
DM components & Milky Way interstellar medium from NE2001 \citep{CordesLazio2002}. Milky Way halo $\mathcal N(55,25^2)$. Intergalactic medium integrated with $f_{\rm IGM}=0.83$ and $f_e=7/8$, with fluctuations following \citet{Macquart2020} with $F=0.31$. Host DM in the rest frame lognormal with median $32.97(1+z)^{0.84}$ and standard deviation of the natural logarithm 1.27 \citep{ZhangGQ2020} \\
Lens galaxies & SIS, with the velocity dispersion function of \citet{Choi2007}. Cross section multiplied by 1.25 \\
Gas in lens galaxies & $n_e(r)=\ADM\eta_g n_0(r/r_0)^{-1.25}$ with $n_0=0.1~\mathrm{cm^{-3}}$, $r_0=10~\mathrm{kpc}$, and $\log_{10}\eta_g\sim\mathcal N(0,0.3^2)$. $\ADM=0.01$ (also 0.001 and 0) \\
General galaxy catalog & Early-type galaxies with $50<\sigma_v<500~\kms$, $0.01<z<2$, and $i<25$, surface density $2.33\times10^4~\mathrm{deg^{-2}}$, all sky. Photometric-redshift error $\sigma_z/(1+z)=0.05$ and velocity-dispersion scatter 0.15~dex. Host inclusion fraction $\fhost=0$ or 1 \\
GGSL catalog & Real entries $6~\mathrm{deg^{-2}}$ and false entries $2.57~\mathrm{deg^{-2}}$ (30\% of all entries), $0.05<z<1.5$. Footprint $20{,}000~\mathrm{deg^2}$. Inclusion probability $\epsilon(\theta_E)=\theta_E^2/(\theta_E^2+0.8''^2)$. Exact redshifts and velocity dispersions \\
\enddata
\tablecomments{All DMs are in units of $\dmunit$. Values introduced by ``also'' refer to the comparison cases.}
\end{deluxetable*}

\subsection{Simulated Radio Survey} \label{sec:sim}

We simulate a five-year all-sky survey with about $5\times10^4$ unlensed events, a fluence threshold of $5~\mathrm{Jy\,ms}$, and an isotropic Gaussian localization error with $\sigpos=1''$, following the coherent all-sky monitor of \citet{Connor:2022bwl}. The cosmology is flat $\Lambda$CDM with $H_0=67.36~\kms\,\mathrm{Mpc^{-1}}$, $\Omega_m=0.3153$, and $\Omega_b=0.0493$ \citep{Planck2018}. FRBs follow the cosmic star formation history, and the fluence of a burst with spectral energy $E_\nu$ is $E_\nu(1+z)^{2-\gamma}/(4\pi d_L^2)$ with $\gamma=1.5$.

Five percent of the sources repeat, comparable to the 3.7\% of the first CHIME/FRB catalog \citep{CHIMEFRB2021Catalog}. Each repeater has $1+\mathrm{Poisson}(3)$ detected bursts at times uniformly distributed over five years, with energies drawn independently above the detection threshold. The DMs of repeat bursts scatter about that of the first burst with standard deviation $\sigrep$, so the DM difference of two later bursts has standard deviation $\sqrt2\,\sigrep$. The burst-to-burst DM scatter of repeaters has not been measured for a population, so we treat it as a parameter, with $\sigrep=1~\dmunit$ for repeaters whose DMs are stable to about $1~\dmunit$ over the survey and $\sigrep=3~\dmunit$ for a less stable population. Secular drifts such as that of FRB 20190520B are not modeled. A repeater then contributes on average 7.5 pairs of bursts, and the numbers of repeater pairs reported below scale with this value.

Lenses are SIS galaxies with the velocity dispersion function of \citet{Choi2007}. The lensing cross section is multiplied by 1.25 to approximate the effect of ellipticity and shear, so that the optical depth differs from the fit of \citet{Oguri:2019fix} by 2\%--11\% for $0.5<z_s<3$. Lensed sources are drawn in proportion to the source redshift distribution times the optical depth, lenses with the cross section as weight, and the source position $y$, in units of $\theta_E$ (Section~\ref{sec:conditions}), with $p(y)=2y$. Lensed fluences are multiplied by $|\mu_\pm|=|1\pm1/y|$, which includes magnification bias. Lenses are drawn from a grid in lens redshift and velocity dispersion, and halving its spacing changes the strict purity and recovery by at most 0.015. Lensed sources repeat like unlensed ones, and all bursts of a lensed repeater share the lens geometry. We do not reject directions with many events as repeaters, because a lensed repeater with $N$ bursts also produces $2N$ events. Pairs that combine events of different bursts of one lensed repeater, which we call cross-burst pairs, are scored as false, because their arrival-time difference is not the lensing delay.

\begin{figure*}
\centering
\includegraphics[width=\textwidth]{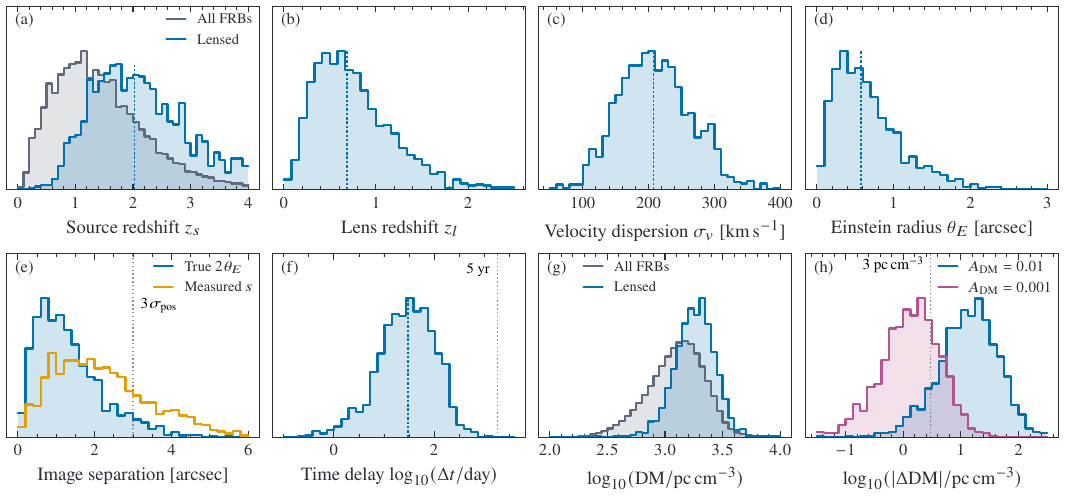}
\caption{Properties of the 1519 recoverable lensed image pairs in the 100 fiducial realizations. All curves are normalized to unit area, and the vertical dashed lines mark the medians of the lensed sample. (a) Source redshifts of all FRBs (gray) and lensed FRBs (blue). (b) Lens redshift. (c) Velocity dispersion of the lens galaxies. (d) Einstein radius. (e) True image separation $2\theta_E$ and measured separation $s$ after adding the localization errors, with the dotted line marking $3\sigpos=3''$. (f) Time delay, with the dotted line marking the five-year survey duration. (g) DMs of all FRBs and lensed FRBs. (h) DM difference between the images for $\ADM=0.01$ (blue) and 0.001 (purple), with the dotted line marking the threshold of $3~\dmunit$. In the gas-poor case most DM differences fall below this threshold.
\label{fig:population}}
\end{figure*}

An image pair is recoverable if the fluences of both images exceed the detection threshold and both arrive within the survey. The 100 fiducial realizations contain 1519 recoverable image pairs, or 15.2 per realization on average, within the 5--40 galaxy-lensed FRBs that \citet{Connor:2022bwl} expect from five years of their array. They belong to 1470 lens systems, because a lensed repeater can contribute several pairs. Figure~\ref{fig:population} shows their properties. The median source redshift is 2.02 (16\%--84\% range 1.29--2.90), higher than the 1.32 of all FRBs because the lensing optical depth grows with source redshift. The median lens redshift is 0.69, the median velocity dispersion $207~\kms$, and the median Einstein radius $0.58''$. The median true separation of $1.15''$ becomes a measured separation of $2.01''$ ($0.96''$--$3.50''$) after the localization errors are added, so the measured separation is set mostly by the localization error. The median delay is 30~days (8--103~days). Lensed image pairs thus share with repeater pairs a separation of order the localization error and an interval of weeks, which is why the DM difference and the optical data are needed to tell them apart.

\subsection{Dispersion Measures and the Source Redshift} \label{sec:dmmodel}

The DM of each event is
\begin{equation}
\mathrm{DM}=\mathrm{DM}_{\rm MW,ISM}+\mathrm{DM}_{\rm MW,halo}+\mathrm{DM}_{\rm IGM}+\frac{\mathrm{DM}_{\rm host,rest}}{1+z}, \label{eq:dm}
\end{equation}
with $\Delta\equiv\mathrm{DM}_{\rm IGM}/\langle\mathrm{DM}_{\rm IGM}\rangle$ following \citet{Macquart2020},
\begin{equation}
p(\Delta\mid z)\propto\Delta^{-3}\exp\left[-\frac{(\Delta^{-3}-C_0)^2}{18\sigma_{\rm DM}^2}\right],\quad\sigma_{\rm DM}=\frac{0.31}{\sqrt{z}}, \label{eq:macquart}
\end{equation}
where $C_0$ ensures $\langle\Delta\rangle=1$ and $z$ is floored at $10^{-3}$. The other components are listed in Table~\ref{tab:model}.

The gas of a lens galaxy follows $n_e(r)=\ADM\eta_gn_0(r/r_0)^{-p}$ with $p=1.25$, where $\eta_g$ is a lognormal galaxy-to-galaxy scatter of 0.3~dex, giving a column density at impact parameter $b$ of
\begin{equation}
N_e(b)=\ADM\eta_g\sqrt{\pi}\,\frac{\Gamma[(p-1)/2]}{\Gamma(p/2)}\,n_0r_0^p\,b^{1-p}. \label{eq:column}
\end{equation}
With $n_0$ in $\mathrm{cm^{-3}}$ and $r_0$, $b$ in kpc, the rest-frame DM is $10^3N_e~\dmunit$, and the observed DM difference between images is the difference of the two values divided by $1+z_l$. The two light paths pass the lens at different impact parameters, so they acquire different DMs. The amplitude $\ADM$ scales the gas-rich model~2 of \citet{Er:2022lad} ($\ADM=1$), and $\ADM=0.01$ gives $n_e\approx10^{-3}~\mathrm{cm^{-3}}$ at 10~kpc, and we treat $\ADM$ as a free parameter. The fiducial $\ADM=0.01$ gives a median $\DDMlens$ of $14~\dmunit$ (16\%--84\% range 4.2--43), within the range of \citet{Er:2022lad}. Gas-poor lenses, with ten times less ionized gas ($\ADM=0.001$), give a median of $1.4~\dmunit$, with three quarters of the image pairs below $3~\dmunit$, and $\ADM=0$ gives vanishing image DM differences. The lens DM common to both images, about $60~\dmunit$, is not added. It would bias $\hat z_s$ high by less than 0.1, below the intrinsic scatter of the DM--redshift relation.

The redshift of an FRB is not measured in the radio catalog. The selection therefore estimates the source redshift from the mean DM of the pair by solving
\begin{equation}
\begin{split}
\overline{\mathrm{DM}}-\mathrm{DM}_{\rm MW,ISM}-55={}&\langle\mathrm{DM}_{\rm IGM}(\hat z_s)\rangle\\
&+32.97(1+\hat z_s)^{-0.16},
\end{split} \label{eq:zhat}
\end{equation}
where the left side subtracts the Milky Way interstellar DM and the mean halo DM of $55~\dmunit$, and the last term is the median host DM in the observer frame. We use the same mean relation that generates the DMs, so $\hat z_s$ scatters about the true redshift only through the fluctuations of the IGM, halo, and host DMs. A bias of the real relation would enter the selection through $\hat z_s$ (Section~\ref{sec:discussion}).

\subsection{Optical Catalogs} \label{sec:catalogs}

The optical catalogs are mock catalogs built for this study and are not tied to the configuration of a specific survey. The general galaxy catalog contains early-type galaxies with $i<25$ over the whole sky, $\Sigma_{\rm gal}=2.33\times10^4~\mathrm{deg^{-2}}$, photometric-redshift errors $\sigma_z/(1+z)=0.05$, and velocity dispersions estimated from the luminosity through the Faber--Jackson relation with 0.15~dex scatter. Variations of depth, completeness, and blending across a real survey are not modeled. Catalog galaxies are drawn from the velocity dispersion function times the comoving volume for $50<\sigma_v<500~\kms$ and $0.01<z<2$, with $L\propto\sigma_v^4$ \citep{FaberJackson1976} normalized to $M_i=-21.5$ at $161~\kms$ and no $K$-correction. The deflectors of 1502 of the 1519 recoverable pairs are in the general catalog, and the other 17 are fainter than $i=25$. The catalog can find lenses with faint hosts but may contain the hosts of unlensed sources. We bracket this with a host inclusion fraction $\fhost=0$ or 1, the probability that the host of an unlensed source at $z<2$ enters the catalog, which absorbs its magnitude and type. A host in the catalog receives the true source redshift, a Gaussian offset of 1~kpc per axis, a velocity dispersion drawn from catalog galaxies at that redshift, and the same measurement errors as other galaxies. Restricting the catalog to early-type galaxies keeps the galaxies that dominate the lensing cross section and excludes most FRB hosts, since about 90\% of known FRB hosts are star-forming \citep{Gordon:2023cgw}, so $\fhost=1$ is pessimistic. Even for $\fhost=1$ the catalog contains the hosts of only about 78\% of the repeater pairs, because hosts beyond its redshift limit of $z=2$ are absent.

A GGSL catalog lists galaxies recognized as lenses because they bend the light of a background galaxy into arcs. The deflector of a lensed FRB enters it when the FRB host is itself lensed into detectable arcs. Our GGSL catalog has the surface density of $6~\mathrm{deg^{-2}}$ forecast for LSST by \citet{Collett:2015roa} for optimal image stacking, which is optimistic, over $20{,}000~\mathrm{deg^2}$. We do not model the brightness of the host and include the deflector of a lensed FRB in this footprint with probability $\epsilon(\theta_E)=\theta_E^2/(\theta_E^2+0.8''^2)$, a parametrization of our own in which lenses with Einstein radii below about $0.8''$, comparable to ground-based seeing, are rarely recognized. We assume a catalog purity of 70\%, so 30\% of the entries are false. Real entries that are not FRB deflectors lens unrelated background galaxies and can produce only chance associations. Redshifts and velocity dispersions of GGSL entries are taken as exact. GGSL velocity dispersions are drawn with weight $\sigma_v^4$, and false entries from the general galaxy distribution. Arc identification is not simulated.

\subsection{Event Pairs and Radio Treatments} \label{sec:veto}

Events within $10''$ of each other are paired, a radius well above the image separations (Figure~\ref{fig:population}(e)). We compare three treatments of repeat bursts before matching, namely no DM-based rejection, a near-duplicate veto, and an all-pair DM cut. The near-duplicate veto treats two events that coincide within three times the larger of their localization errors, $s<3\max(\sigma_1,\sigma_2)$, and have $\DDM<3~\dmunit$ as repeated detections of one source, and keeps only the earlier one. Its DM limit equals $3\sigrep$ of the fiducial model, so the veto is itself set for small $\sigrep$. It removes most repeat bursts but also discards image pairs that happen to have a small separation and a small DM difference (Section~\ref{sec:cost}). The all-pair DM cut additionally removes every pair with $\DDM<T$ at any separation within $10''$, with $T=3~\dmunit$ unless stated otherwise (Section~\ref{sec:dm} scans $T$ up to $20~\dmunit$).

For the radio-only reference, we use another 100 independent realizations to set a calibrated window containing 99\% of the recoverable image pairs, $s<5.61''$, $\Delta t<338$~days, and $\DDM<153~\dmunit$. Applied after the near-duplicate veto it gives the calibrated radio selection, and adding the all-pair DM cut gives the radio-only DM selection, our reference. Per realization, these two selections reduce the 344.6 event pairs of Section~\ref{sec:background} to 166.1 and 36.1, whereas without the veto the calibrated window contains 5505. The optical schemes do not use this window. They start from all pairs within $10''$ that survive the radio treatment and bound the delay physically through Equation~(\ref{eq:dtmax}). Because simulated burst times are uniform, the window's limit $\Delta t<338$~days removes repeater pairs that real repeaters, which are clustered in time, would not lose, so the radio reference is optimistic.

\subsection{Optical Conditions} \label{sec:conditions}

We adopt the singular isothermal sphere (SIS) lens \citep{NarayanBartelmann1996}. For velocity dispersion $\sigma_v$, the Einstein radius is $\theta_E=4\pi(\sigma_v/c)^2D_{ls}/D_s$, and for a source at $y\equiv\beta/\theta_E<1$ two images form at $(1\pm y)\theta_E$ on opposite sides of the lens, separated by $2\theta_E$, with a delay
\begin{equation}
\Delta t=\frac{2y\,\theta_E^2D_{\Delta t}}{c},\qquad D_{\Delta t}\equiv(1+z_l)\frac{D_lD_s}{D_{ls}}, \label{eq:delay}
\end{equation}
where $z_l$ and $D_l$ are the lens redshift and distance.

For each event pair, we search the optical catalog and apply five conditions and a mass pre-screen in turn. The first is a position condition. In coordinates along and perpendicular to the line joining the events, with origin at their midpoint, the galaxy offsets $u$ and $v$ must satisfy
\begin{equation}
|v|<3\sigpos,\qquad|u|<s/2+3\sigpos, \label{eq:position}
\end{equation}
where $s$ is the measured separation and $\sigpos$ the per-axis localization error. A mass pre-screen then requires the maximum Einstein radius of the galaxy, for a source at $z_s=4$, to exceed $0.3\,s/2$. The second condition requires $z_l<\hat z_s$, with $\hat z_s$ from Equation~(\ref{eq:zhat}), since a lens must lie in front of the source. The third condition uses $y<1$ to bound the delay. With the observed Einstein radius $\thEobs\equiv(r_A+r_B)/2$ defined by the distances $r_A$ and $r_B$ of the galaxy to the two events, which for an SIS equals $\theta_E$ because the images lie at $(1\pm y)\theta_E$,
\begin{equation}
\Delta t_{\rm obs}<1.2\,\frac{2\thEobs^2D_{\Delta t}(z_l,\hat z_s)}{c}, \label{eq:dtmax}
\end{equation}
where the factor 1.2 leaves a 20\% margin for the errors of $\thEobs$ and $\hat z_s$. The fourth condition requires the Einstein radius inferred from $\sigma_v$ and the redshifts to match,
\begin{equation}
|\theta_E(\sigma_v,z_l,\hat z_s)-\thEobs|<0.7\,\thEobs. \label{eq:thetamatch}
\end{equation}
Because $\thEobs\ge s/2$, a galaxy whose maximum Einstein radius for $z_s=4$ is below $0.3\,s/2$ cannot satisfy Equation~(\ref{eq:thetamatch}) for any $\hat z_s\le4$, so the mass pre-screen only discards such galaxies early. The fifth condition, $\DDM<150~\dmunit$, rejects pairs of unrelated sources, whose DMs typically differ by hundreds of $\dmunit$, while keeping more than 99\% of the image pairs (compare the calibrated limit of $153~\dmunit$ in Section~\ref{sec:veto}). Because such chance pairs are rare, removing it leaves the false candidates and the recovery essentially unchanged (Section~\ref{sec:cost}). The tolerance of 0.7 in Equation~(\ref{eq:thetamatch}) was chosen on realizations independent of the evaluation sample as a compromise between purity and recovery, and Section~\ref{sec:cost} shows the full trade-off. The time-delay and DM conditions hardly affect the results, so their margins are not critical. Every pair--galaxy combination that passes all conditions is kept as a candidate. When several galaxies pass for one event pair, no single deflector is chosen.

\subsection{Evaluation} \label{sec:eval}

A pair is correct if its events are the two images of one burst. The pair purity is the fraction of correct pairs among the distinct event pairs in the output. The strict purity of the optical schemes counts candidates, that is pair--galaxy combinations, and additionally requires the galaxy to be the true deflector. The recovery fraction is the fraction of recoverable image pairs that are recovered, with the correct deflector for optical schemes. Results are averaged over 100 realizations, with 95\% intervals from 10,000 cluster bootstrap resamplings over realizations. Because the event rate, exposure fraction, and lensing cross section rescale the numbers of pairs but not the probability that a given pair passes, the recovery fraction and the reduction factors, the ratios of false pairs before and after optical matching, measure the discriminating power. Purities instead depend on the ratio of repeater to image pairs. For a repeater background $k$ times larger, for example through a higher repeater fraction or more bursts per repeater, a purity $P$ becomes $P/[P+k(1-P)]$.

\section{Results} \label{sec:results}

\subsection{Radio-only Selection and the DM Threshold} \label{sec:dm}

\begin{figure*}
\centering
\includegraphics[width=\textwidth]{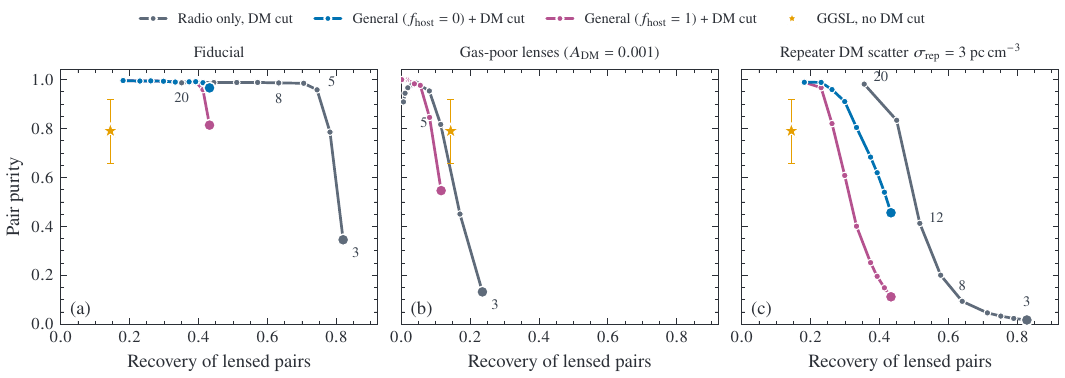}
\caption{Pair purity against recovery as the threshold $T$ of the all-pair DM cut increases from 3 to $20~\dmunit$ (points at 3, 4, 5, 6, 8, 10, 12, 15, and $20~\dmunit$, some labeled, with the large marker at $T=3~\dmunit$), for the radio-only selection (gray) and the general galaxy catalog after the same cut with $\fhost=0$ (blue) and 1 (purple). The orange star is GGSL matching without DM-based rejection, with its 95\% cluster bootstrap interval. (a) Fiducial model, $\sigrep=1~\dmunit$. (b) Gas-poor lenses ($\ADM=0.001$, $\fhost=1$ only). (c) $\sigrep=3~\dmunit$. Optical recovery requires the correct deflector. Intervals of the curves are omitted for clarity.
\label{fig:dmthr}}
\end{figure*}

Figure~\ref{fig:dmthr} shows pair purity against recovery as the threshold $T$ of the all-pair DM cut increases from 3 to $20~\dmunit$. For the radio-only selection (gray), the outcome is set by the relation among $T$, $\sigrep$, and $\DDMlens$ (Section~\ref{sec:background}). In the fiducial model with $\sigrep=1~\dmunit$, the repeater pairs surviving $T=3~\dmunit$ lie in the Gaussian tail, and the pair purity rises from 0.345 at $T=3~\dmunit$ to 0.958 at a recovery of 0.744 for $T=5~\dmunit$ and 0.985 at 0.705 for $T=6~\dmunit$.

A threshold tuned to this model, however, fails when $\sigrep$ or $\DDMlens$ differs from the assumed values. For $\sigrep=3~\dmunit$, the cut at $T=5~\dmunit$ gives a pair purity of only 0.033, and a purity above 0.8 requires $T=15~\dmunit$, with a recovery of 0.45. A threshold of $20~\dmunit$ gives a purity of about 0.98 in both models but recovers only 0.35 of the image pairs. For gas-poor lenses, any DM cut removes most image pairs, leaving a recovery of 0.236 at $T=3~\dmunit$, 0.115 at $T=5~\dmunit$, and 0.007 at $T=20~\dmunit$, and for zero DM difference no image pair survives. No single threshold therefore serves the three DM conditions. The cut at $T=5~\dmunit$ fails for $\sigrep=3~\dmunit$, and $T=20~\dmunit$ fails for gas-poor lenses. The DM drift of FRB 20190520B would accumulate over a five-year survey to more than the median $\DDMlens$ of $14~\dmunit$, so no threshold separates such a repeater from a lens. A threshold chosen without knowledge of $\sigrep$ and $\DDMlens$ therefore either keeps repeater pairs or removes image pairs.

\subsection{Optical Criterion at a Fixed DM Threshold} \label{sec:main}

We now fix $T=3~\dmunit$, the lowest threshold scanned and equal to the DM limit of the near-duplicate veto, and add the optical catalogs (Table~\ref{tab:results}). This threshold keeps the most image pairs, including a quarter of those of gas-poor lenses, but suits only small $\sigrep$. Relative to the radio-only DM selection, the general galaxy catalog reduces the false pairs from 23.61 to 0.23 and 1.51 per realization for $\fhost=0$ and 1, raising the pair purity from 0.345 to 0.966 and 0.814. The GGSL catalog leaves 2 false pairs in 100 realizations, a one-sided 95\% Poisson upper limit of 0.063 per realization, and a purity of 0.990. The cost is recovery, 0.432 and 0.124 against 0.820.

\begin{deluxetable*}{llrrcc}
\tablecaption{Results for Three Radio Treatments and Four Selection Schemes \label{tab:results}}
\tablehead{\colhead{Radio treatment} & \colhead{Selection scheme} & \colhead{Correct pairs} & \colhead{False pairs} & \colhead{Pair purity} & \colhead{Recovery fraction}}
\startdata
No DM-based rejection & Radio only (calibrated window) & 14.73 & 5490.67 & 0.003 [0.003, 0.003] & 0.970 [0.960, 0.978] \\
 & General galaxies, $\fhost=0$ & 7.68 & 60.64 & 0.112 [0.103, 0.123] & 0.502 [0.475, 0.529] \\
 & General galaxies, $\fhost=1$ & 7.68 & 431.58 & 0.017 [0.016, 0.019] & 0.502 [0.475, 0.529] \\
 & GGSL & 2.19 & 0.58 & 0.791 [0.659, 0.920] & 0.144 [0.127, 0.162] \\
Near-duplicate veto & Radio only (calibrated window) & 12.85 & 153.24 & 0.077 [0.072, 0.082] & 0.846 [0.828, 0.863] \\
 & General galaxies, $\fhost=0$ & 6.79 & 1.24 & 0.846 [0.818, 0.871] & 0.444 [0.418, 0.470] \\
 & General galaxies, $\fhost=1$ & 6.79 & 7.11 & 0.488 [0.462, 0.515] & 0.444 [0.418, 0.470] \\
 & GGSL & 1.96 & 0.04 & 0.980 [0.953, 1.000] & 0.129 [0.113, 0.145] \\
All-pair DM cut ($T=3$) & Radio only (calibrated window) & 12.45 & 23.61 & 0.345 [0.326, 0.365] & 0.820 [0.801, 0.837] \\
 & General galaxies, $\fhost=0$ & 6.61 & 0.23 & 0.966 [0.951, 0.979] & 0.432 [0.407, 0.457] \\
 & General galaxies, $\fhost=1$ & 6.61 & 1.51 & 0.814 [0.784, 0.842] & 0.432 [0.407, 0.457] \\
 & GGSL & 1.89 & 0.02 & 0.990 [0.973, 1.000] & 0.124 [0.109, 0.141] \\
\enddata
\tablecomments{Numbers of pairs are averages per all-sky realization, and brackets give 95\% cluster bootstrap intervals. A correct pair does not require the correct deflector, whereas the recovery fraction of the optical schemes does, so for the general galaxy catalog the correct pairs divided by the 15.19 recoverable pairs slightly exceed the recovery fraction. For the GGSL rows with DM-based rejection, the false pairs occur in only a few realizations and the upper ends of the intervals are not meaningful.}
\end{deluxetable*}

The optical criterion removes false pairs selectively. Relative to the radio-only DM selection, the general galaxy catalog outputs 0.53 times as many correct pairs but only 0.064 and 0.010 times as many false pairs for $\fhost=1$ and 0. The GGSL catalog outputs 0.15 times as many correct pairs and less than 0.003 times as many false pairs. None of the 2347 repeater pairs in 100 realizations survives GGSL matching, a reduction by at least 780 given the 95\% upper limit of three events, and both remaining GGSL false pairs are cross-burst pairs. These limits reflect the number of realizations, since only about $10^{-3}$ chance GGSL entries per realization are expected within the search windows of the surviving repeater pairs.

In the fiducial model, a tuned radio cut ($T=5$--$6~\dmunit$) matches the purity of the optical schemes at a higher recovery, so the optical criterion adds little there. Its gain appears where the threshold does not match the DM conditions. For $\sigrep=3~\dmunit$ and $T=3~\dmunit$, the purity rises from 0.017 to 0.456 with the general galaxy catalog ($\fhost=0$, or 0.112 for $\fhost=1$) and to 0.927 with the GGSL catalog. For gas-poor lenses at the same threshold, it rises from 0.132 to 0.903 ($\fhost=0$, or 0.546 for $\fhost=1$), with the recovery falling from 0.236 to 0.117. GGSL matching without DM-based rejection does not depend on the DM conditions. In the fiducial case, for gas-poor lenses, for zero DM differences, and for $\sigrep=3~\dmunit$, its recovery fraction is 0.144 and its pair purity 0.79 (0.66--0.92). Of its 0.58 false pairs per realization, 0.42 are cross-burst pairs and 0.16 are repeater pairs with unrelated GGSL entries. Because cross-burst pairs point to genuine lenses, the purity for identifying lens systems is about 0.94. Without DM-based rejection, the general galaxy catalog with $\fhost=1$ leaves 432 false pairs per realization, so only the GGSL catalog can be used alone. For $\sigrep=3~\dmunit$, the conservative radio cut at $T=20~\dmunit$ (pair purity 0.982, recovery 0.355) does better than the general galaxy catalog at $T=3~\dmunit$, and adding the general catalog ($\fhost=0$) at $T=20~\dmunit$ raises the purity only to 0.989 while halving the recovery. The general catalog therefore helps mainly when the threshold is set too low, whereas GGSL matching without DM-based rejection removes the need to choose a threshold. Because it uses no DM-based rejection, it is also unaffected by secular DM drifts of repeaters, except through $\hat z_s$ and the condition $\DDM<150~\dmunit$.

\subsection{Why Repeater Pairs Are Excluded} \label{sec:mechanism}

\begin{figure*}
\centering
\includegraphics[width=0.85\textwidth]{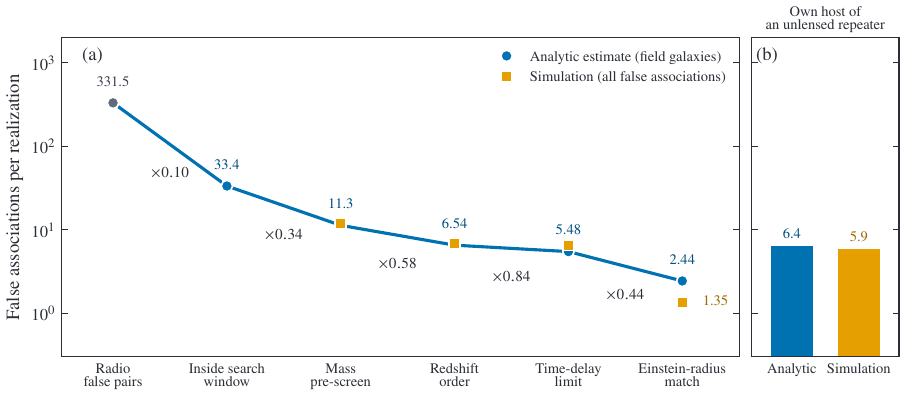}
\caption{Progressive reduction of false associations by the optical conditions (general galaxy catalog, near-duplicate veto). (a) False associations remaining after each condition, starting from 331.5 false pairs per realization (gray point). Blue points are the analytic estimate for unrelated galaxies (Equation~\ref{eq:ngal}), with the fraction retained at each step. Orange squares are all false associations in the simulation, recorded from the mass pre-screen onward, dominated by unrelated galaxies with a few cross-burst pairs. (b) Misidentifications of the hosts of unlensed repeaters for $\fhost=1$.
\label{fig:mechanism}}
\end{figure*}

Sections~\ref{sec:mechanism}--\ref{sec:localization} use the near-duplicate veto, which isolates the optical conditions from the choice of $T$, and report strict purities, which in the fiducial case are 0.833 and 0.482 for $\fhost=0$ and 1. A false association is a candidate, that is a pair--galaxy combination, that is not a genuine image pair with its true deflector. Unrelated galaxies and repeater hosts are excluded in different ways (Figure~\ref{fig:mechanism}). For unrelated galaxies, Equation~(\ref{eq:ngal}) gives 33.4 galaxies in the search windows of the 331.5 false pairs per realization under the near-duplicate veto, and the mass pre-screen and the redshift, time-delay, and Einstein-radius conditions reduce them analytically to 2.44. For this estimate, each condition is applied, for every false pair with its measured $s$, $\Delta t$, and $\hat z_s$ and with $\thEobs=s/2$, to galaxies drawn from the redshift and velocity-dispersion distribution of the catalog, and the factors in Figure~\ref{fig:mechanism}(a) are the resulting mean fractions. The simulation leaves 1.35 false associations, 1.26 of them with unrelated galaxies and the rest cross-burst pairs, in agreement with the estimate within a factor of two. A host in the catalog, in contrast, is not suppressed by the small window. In each realization the hosts of about 260 repeater pairs are in the catalog for $\fhost=1$, 95 pass the position condition and the mass pre-screen, which rejects low-mass hosts, the redshift condition removes about two thirds of them because $\hat z_s$ lies below the photometric redshift of the host, and only 5.9 survive the time-delay and Einstein-radius conditions because $D_{ls}\ll D_s$. The same analytic estimate, averaged over the redshift and velocity-dispersion distribution of hosts, gives 6.4 (Figure~\ref{fig:mechanism}(b)).

\subsection{The Cost in Genuine Image Pairs} \label{sec:cost}

The optical conditions also remove genuine image pairs, and the Einstein-radius condition removes the largest share. Of the 1519 recoverable image pairs, the near-duplicate veto removes 188, and the position condition and the mass pre-screen remove a further 245, including the 17 pairs whose deflectors are not in the catalog. After the redshift and time-delay conditions 1053 pairs remain, after the Einstein-radius condition 678, and 674 are in the final output. The loss comes from the errors of the inputs. The velocity dispersion from the luminosity has a scatter of 0.15~dex, and since $\theta_E\propto\sigma_v^2$ the inferred Einstein radius scatters by about 0.3~dex. Together with the localization errors and the error of $\hat z_s$, this places some genuine pairs outside the tolerance of Equation~(\ref{eq:thetamatch}).

\begin{figure*}
\centering
\includegraphics[width=\textwidth]{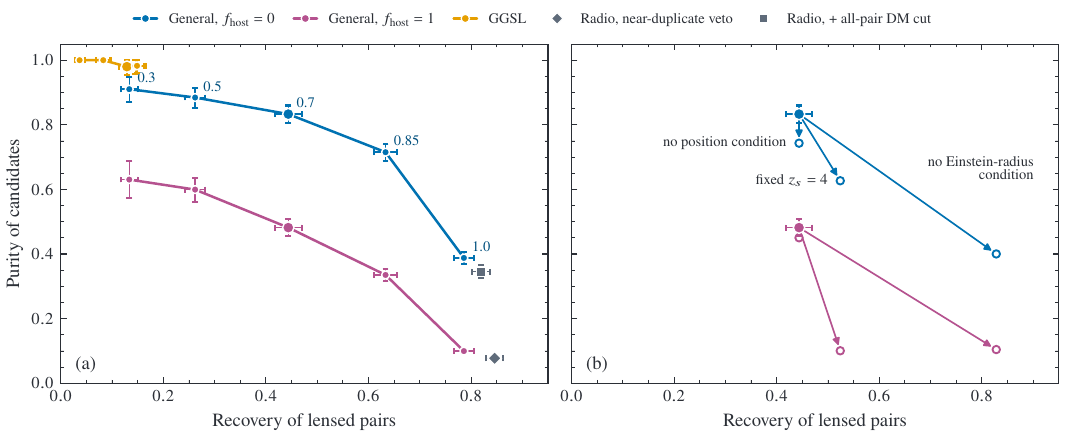}
\caption{Purity--recovery trade-off (near-duplicate veto). (a) Strict purity and recovery of the optical schemes as the Einstein-radius tolerance varies from 0.3 to 1.0, with large circles at the fiducial 0.7 and the tolerance values labeled for $\fhost=0$. The gray diamond and square show the calibrated radio selection and the radio-only DM selection ($T=3~\dmunit$), for which the vertical axis is the pair purity. (b) Displacement of the general galaxy catalog when one condition is removed ($\fhost=0$ blue, $\fhost=1$ purple). Removing the time-delay or DM condition changes the purity by at most 0.033 and is not shown. Error bars are 95\% cluster bootstrap intervals.
\label{fig:tradeoff}}
\end{figure*}

Relaxing the tolerance trades purity for recovery (Figure~\ref{fig:tradeoff}(a)). For the general galaxy catalog with $\fhost=0$, relaxing it from 0.3 to 1.0 raises the recovery from 0.134 to 0.786 and lowers the strict purity from 0.910 to 0.388. At a tolerance of 1.0 the recovery approaches that of the calibrated radio selection (0.846), while the purity is about 5 times the radio value of 0.077. For $\fhost=1$ the same relaxation gives a purity of only 0.100, so with many hosts in the catalog a high purity requires a tight tolerance and a low recovery. The GGSL strict purity is at least 0.98 over the whole range, with recovery between 0.037 and 0.149. The low GGSL recovery is set mainly before matching. Under the near-duplicate veto it factorizes as $R_{\rm GGSL}=(746/1519)\times(265/746)\times(196/265)\approx0.491\times0.355\times0.740=0.129$, the product of the footprint, the inclusion probability, and the efficiency of the criterion, so the first two factors cap the recovery at 0.174. Figure~\ref{fig:tradeoff}(b) shows the effect of removing one condition at a time. Without the Einstein-radius condition, the strict purity of the general galaxy catalog drops from 0.833 to 0.400 for $\fhost=0$ and from 0.482 to 0.104 for $\fhost=1$, with recovery rising from 0.444 to 0.828. Replacing $\hat z_s$ by a fixed $z_s=4$ increases the false candidates for $\fhost=1$ from 7.25 to 71.27 per realization. Removing the position condition lowers the strict purity for $\fhost=0$ only from 0.833 to 0.743, because galaxies off the line joining the events have larger $\thEobs$ and are mostly rejected by the Einstein-radius condition, while removing the time-delay or DM condition changes it by at most 0.033.

\subsection{Dependence on Localization Precision} \label{sec:localization}

\begin{figure*}
\centering
\includegraphics[width=0.9\textwidth]{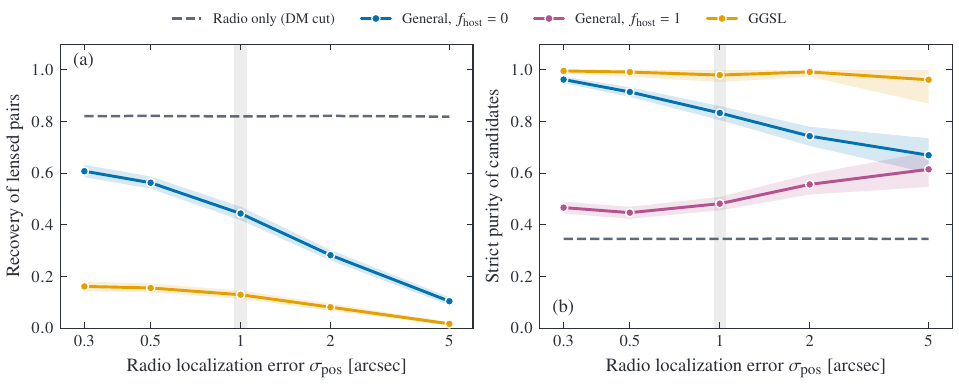}
\caption{Effect of the per-axis radio localization error (near-duplicate veto). (a) Recovery fraction, which for the general galaxy catalog does not depend on $\fhost$. (b) Strict purity of the optical schemes, with the pair purity of the radio-only DM selection as the gray dashed line. Bands are 95\% cluster bootstrap intervals, and the gray vertical band marks the fiducial $\sigpos=1''$.
\label{fig:localization}}
\end{figure*}

The radio-only DM selection hardly changes for $0.3''\le\sigpos\le5''$ because the near-duplicate radius scales with $\sigpos$, whereas the optical criterion depends strongly on it (Figure~\ref{fig:localization}). From $\sigpos=1''$ to $0.3''$, the general galaxy catalog improves in recovery from 0.444 to 0.608 and in strict purity for $\fhost=0$ from 0.833 to 0.962, and the GGSL recovery rises from 0.129 to 0.161. At $5''$ the recovery falls to 0.104 for the general galaxy catalog and 0.016 for the GGSL catalog, while the GGSL strict purity stays above 0.96.

\section{Discussion} \label{sec:discussion}

Rejecting every pair without an optical match would discard the genuine pairs that the optical criterion misses, so the criterion is better used to grade candidates. Ranking GGSL matches without DM rejection first, general galaxy matches after the DM cut second, and the remaining pairs of the radio-only DM selection third, the first two tiers recover 0.487 of the image pairs with a pair purity of 0.90 in the fiducial case (0.78 for $\fhost=1$). All three tiers recover 0.852 at a pair purity of 0.35, more than radio only (0.820 at 0.345), because the first tier also contains image pairs that the DM cut removes. For gas-poor lenses and $\fhost=0$, the first two tiers recover 0.237 at a pair purity of 0.82, about the recovery of radio only at 0.13.

Whatever the tier, candidates must be confirmed by follow-up observations. In the radio band, the polarization, rotation measure, and high-time-resolution dynamic spectra of the two images can be compared. In the optical and infrared bands, the mass distribution of the deflector can be measured, lensed images of the host can be searched for, and the spectroscopic redshift of the host can be determined. For lensed repeaters, all bursts give the same delay, which confirms the lensing geometrically \citep{Wucknitz:2020spz}. Only after a candidate is confirmed and its lens mass model is constrained can its delay be used to measure $H_0$ \citep{Li:2017mek,Wucknitz:2020spz,Zhang:2024rra}.

The optical criterion is complementary to radio morphology. Lensed images share the intrinsic time--frequency structure of one burst, so the similarity of dynamic spectra can also reject repeater pairs \citep{Chang:2024hxi}, although scattering and multipath propagation complicate it \citep{Kader:2025tia}. The two use independent information and can be combined, although a morphology-augmented radio reference would reduce the gains quoted here.

The events are distributed over the whole sky, whereas the array of \citet{Connor:2022bwl} monitors $5000~\mathrm{deg^2}$ continuously, so its event density is about 8 times higher. Chance pairs remain negligible even after this factor, and the optical criterion depends only on galaxies near each pair. Because that array monitors continuously, it also records second images with delays of up to years, as our simulation does. For transit surveys, the number of image pairs at fixed event number scales with the exposure fraction while repeater pairs do not, which changes purities but not the reduction factors.

Our results rest on idealizations. The SIS model admits only doubles, and ellipticity and shear will scatter the relation between separation and Einstein radius, so the tolerances of Equations~(\ref{eq:dtmax}) and (\ref{eq:thetamatch}) should be revisited with realistic lenses. The GGSL parameters are exact and its inclusion ignores whether a faint host at $z_s\approx2$ forms detectable arcs, which is optimistic for its purity and recovery. An all-type catalog would raise unrelated associations in proportion to $\Sigma_{\rm gal}$, whereas a clean early-type selection excludes most star-forming hosts. Since host associations add linearly, $\fhost=0.1$ would give about 0.36 false pairs per realization under the all-pair DM cut at $T=3~\dmunit$, a reduction by about 65 and a pair purity of about 0.95. The source redshift is estimated with the same DM--redshift relation that generates the DMs, so a bias in this relation, uncalibrated at $z_s\approx2$, would shift the redshift and Einstein-radius conditions, and catastrophic photometric-redshift outliers and DM measurement errors are not included. Hosts are offset by 1~kpc per axis, less than the observed median of 3.3~kpc, but both lie well within the $3''$ search half-width beyond $z\approx0.1$. Finally, the method requires localization of about one arcsecond or better, as provided by interferometric arrays such as CASM-256, DSA/Chronoscope, BURSTT, and the SKA \citep{Connor:2026uxo,Connor:2026nex,Lin2022BURSTT,Lin2025arXiv251207300,Macquart2015SKA}. The FAST Core Array, which adds 24 antennas of 40~m within 5~km of FAST, will provide a $4.3''$ beam at 1.4~GHz \citep{Jiang:2024fca}. Bright bursts can be localized to a fraction of this beam, and the sensitivity of FAST helps to detect the fainter, demagnified image of a lensed pair, which must also exceed the detection threshold for a pair to be recoverable.

\section{Conclusions} \label{sec:conclusions}

Unlensed repeaters dominate the background for identifying strongly lensed FRBs. A radio-only DM cut is effective only if $\sigrep\ll T\ll\DDMlens$, but both quantities are uncertain, and at least one observed repeater drifts in DM by more than a typical $\DDMlens$ over a survey. A threshold tuned to one repeater DM variation fails for a larger one, and a conservative threshold loses the image pairs of gas-poor lenses. Requiring a foreground galaxy whose position, redshift, mass, and maximum time delay match the pair provides a criterion based on independent optical information. At a fixed DM cut it raises the purity of the candidates from 0.35 to 0.97--0.99 in the fiducial model and from 0.02 to 0.46--0.93 for a three times larger repeater DM variation, provided for the general catalog that repeater hosts are largely absent from it, while keeping about half (general catalog) or one seventh (GGSL catalog) of the image pairs that the cut retains. In the fiducial model a better-tuned DM cut reaches a similar purity at a higher recovery. An optimistic LSST-like GGSL catalog without DM-based rejection of repeaters gives a purity of about 0.79 at a recovery of 0.14 whatever the DM conditions, including vanishing image DM differences, for which no all-pair DM cut recovers any image pair. Optical catalogs can therefore be used to identify strongly lensed FRBs in the arcsecond-localized samples that forthcoming arrays will provide, and the candidates, once graded by the reliability of their optical matches and confirmed by follow-up, can serve time-delay cosmology.

\begin{acknowledgments}
We thank Yun Chen for helpful discussions. This work was supported by the National Natural Science Foundation of China (Grant Nos. 12675064, 12533001, 12575049, 12473001, 12305068, and 12305069), the National SKA Program of China (Grant Nos. 2022SKA0110200 and 2022SKA0110203), the China Manned Space Program (Grant No. CMS-CSST-2025-A02), and the 111 Project (Grant No. B16009). Anthropic's Claude was used to improve the language of the manuscript and to assist with code modification and numerical checks. The authors take full responsibility for the content of this work. The simulation, selection, and scoring codes, the numerical results, and the plotting scripts will be made publicly available, and the repository address and archive DOI will be given in the published version.
\end{acknowledgments}

\software{NumPy \citep{Harris:2020xlr}, SciPy \citep{Virtanen:2019joe}, Astropy \citep{Astropy:2013muo,Astropy:2018wqo,Astropy:2022ucr}, pandas \citep{McKinney2010pandas}, Matplotlib \citep{Hunter2007Matplotlib}, NE2001 \citep{CordesLazio2002}}

\bibliography{refs}{}
\bibliographystyle{aasjournalv7.1}

\end{document}